\documentclass{article}

\usepackage{arxiv}

\usepackage[utf8]{inputenc}
\usepackage{float}
\usepackage{graphicx}
\usepackage{subcaption}
\usepackage{natbib}
\usepackage{authblk}
\usepackage{footnote}
\usepackage{url}
\usepackage[acronym,nogroupskip,nopostdot]{glossaries}
\usepackage{booktabs}
\usepackage{xltabular}
\usepackage{multirow, multicol, makecell, booktabs}
\usepackage{amsmath}
\usepackage{amsfonts}
\usepackage{optidef}
\usepackage{subfiles}
\usepackage{nameref}
\usepackage{hyperref}
\usepackage{adjustbox}
\usepackage{indentfirst}
\usepackage{array,ragged2e}
\usepackage{pdflscape}
\usepackage{longtable}

\usepackage{tikz}
\usetikzlibrary{shapes, arrows, positioning}

\usepackage{cleveref}
\usepackage{geometry}
\graphicspath{ {./images/} }

\usepackage{pgf-pie}

\usepackage{setspace}

\setcitestyle{authoryear,open={(},close={)}} 

\title{High School Summer Camps Help Democratize\\ Coding, Data Science, and Deep Learning}

\date{}

\author{Rosemarie~Santa González}
\author{Tsion Fitsum}
\author{Michael Butros}

\affil{H. Milton Steward School of Industrial and Systems Engineering, \protect\\ Georgia Institute of Technology}

\begin{document}

\maketitle

\begin{abstract}
\noindent
This study documents the impact of a summer camp series that introduces high school students to coding, data science, and deep learning. Hosted on-campus, the camps provide an immersive university experience, fostering technical skills, collaboration, and inspiration through interactions with mentors and faculty. Campers' experiences are documented through interviews and pre- and post-camp surveys. Key lessons include the importance of personalized feedback, diverse mentorship, and structured collaboration. Survey data reveals increased confidence in coding, with 68.6\% expressing interest in AI and data science careers. The camps also play a crucial role in addressing disparities in STEM education for underrepresented minorities. These findings underscore the value of such initiatives in shaping future technology education and promoting diversity in STEM fields.
\end{abstract}

\section{Introduction} \label{SecIn}
The importance of high school students learning to code is critical in today’s technology-driven world. As industries increasingly rely on automation, artificial intelligence, and data analytics, the demand for technical skills is growing rapidly. Employment in jobs requiring these skills is projected to see the third-largest increase from 2021 to 2031 \citep{BureauStatistic_2023}. However, despite this rising demand, only 57.5\% of public high schools in the United States offer foundational computer science courses, with significant disparities in access across different regions and demographics \citep{Codeorg_2023}. This lack of access is particularly concerning as it limits opportunities for many students to develop the skills necessary for future success in the workforce. Moreover, enrollment in computer science courses remains alarmingly low, with just 5.8\% of students across 35 states participating in these essential courses \citep{Codeorg_2023}. The situation is even more challenging for minority students, who are 1.4 times less likely than their white and Asian peers to enroll in foundational computer science, even when such courses are available at their schools \citep{Codeorg_2023}. These disparities highlight the urgent need for targeted educational interventions that can bridge the gap and provide all students with the opportunity to acquire critical technical skills.

In response to these disparities, data science summer camps were established to offer high school students essential coding and data science education in an engaging and accessible format. These week-long programs are designed to expose students to coding, data science, and deep learning through a structured, progressive learning experience that spans three levels during the summer. By introducing students to these fields in a supportive environment, the camps aim to spark curiosity and enthusiasm for computer science, allowing participants to build confidence and competence as they progress at their own pace. The significance of these camps extends beyond merely teaching technical skills; they are also a crucial tool in promoting equity in STEM education. By providing a platform where students from diverse backgrounds can learn and grow together, these camps help to address the underrepresentation of minorities in computer science and related fields. Through this study, we document the lessons learned from the 2024 summer camps, focusing on what worked best from the coordinators' perspectives. Additionally, we analyze interviews with student participants to further illustrate the need for and impact of these programs, emphasizing their role in shaping the next generation of tech-savvy, innovative thinkers.

\section{Literature Review} 
\label{LitRev} 
The literature underscores the significant impact of early STEM exposure on students' academic and career paths. For instance, students engaged in STEM during high school are more likely to pursue and complete degrees in these fields \citep{maltese2011pipeline}. Hands-on, experiential learning, as described by Kolb's Experiential Learning Theory \citep{kolb2014experiential}, enhances understanding and retention of complex concepts, which is effectively applied in data science summer camps through real-world problem-solving. K-12 STEM camps are crucial in shaping the academic trajectories of minority students by providing access to advanced fields like AI and machine learning, along with essential mentorship \citep{kong2014association, foltz2014factors}. Programs like the Data Science for Social Good \citealp{sabatti2022near} demonstrate how near-peer mentoring can inspire and support underrepresented groups. Early exposure to programming equips students with critical thinking and problem-solving skills, fostering interest in STEM and preparing them for a tech-driven job market \citep{goode2010connecting, margolis2002unlocking, wilson2010running}. Additionally, these experiences promote creativity, collaboration, and equity, empowering students from diverse backgrounds to succeed in the tech industry \citep{kafai2014connected}. Summer coding camps also have a lasting impact on career paths by boosting students' confidence and interest in technology fields \citep{ncwit2020engaging}. These camps often lead to advanced studies and career opportunities, while addressing gender and diversity gaps in tech \citep{gwc2021impact}. By emphasizing both technical skills and social relevance, these programs foster a commitment to using technology for social good, particularly among minority students \citep{visser2016computer, yilmaz2009hands, mccullough2002attracting}.

Our study contributes to this body of work by analyzing how summer camps effectively combine skill development with social impact, promoting diversity and reducing educational disparities. This documentation is vital for guiding future initiatives that aim to democratize access to technology education and prepare a more diverse workforce.

\section{Data Science Camps} \label{Sec:SBC}
The camps are week-long Computational and Data Science programs for high school students, divided into three levels focusing on coding, data science, and deep learning. These camps provide foundational knowledge for students, especially those with little or no prior experience, preparing them for careers in computational science and related fields. Hosting the camps on-campus allows students to experience the university environment while engaging with advanced technology and methodologies.

\subsection{Program Goals and Structure}
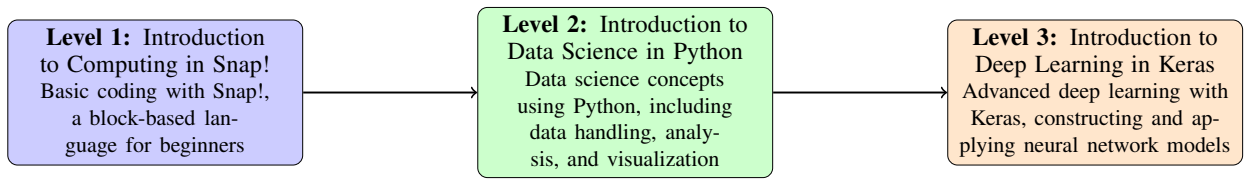
\begin{figure}[h!]
    \centering
    \resizebox{\textwidth}{!}{
    \begin{tikzpicture}[
        node distance=2.5cm and 2.5cm,
        every node/.style={rectangle, rounded corners, text centered, draw=black},
        level1/.style={rectangle, draw=black, fill=blue!20, text width=4cm, minimum height=1cm},
        level2/.style={rectangle, draw=black, fill=green!20, text width=4cm, minimum height=1cm},
        level3/.style={rectangle, draw=black, fill=orange!20, text width=4cm, minimum height=1cm},
        ]

        \node (level1) [level1] {\textbf{Level 1:} Introduction to Computing in Snap!\\ \small{Basic coding with Snap!, a block-based language for beginners}};
        \node (level2) [level2, right=of level1] {\textbf{Level 2:} Introduction to Data Science in Python\\ \small{Data science concepts using Python, including data handling, analysis, and visualization}};
        \node (level3) [level3, right=of level2] {\textbf{Level 3: }Introduction to Deep Learning in Keras\\ \small{Advanced deep learning with Keras, constructing and applying neural network models}};

        \draw[->, thick] (level1) -- (level2);
        \draw[->, thick] (level2) -- (level3);

    \end{tikzpicture}}
    \caption{Progression through the levels of the program} \label{Fig.Levels}
    \label{fig:levels}
\end{figure}

The primary goal of the camps is to introduce high school students to computer science and related career opportunities. These camps are designed for beginners, requiring only basic web navigation skills and an interest in computational and data science. The levels build upon each other, with Level 1 as a prerequisite for Level 2, and Level 2 for Level 3. The structure of the program is visually represented in Figure \ref{Fig.Levels}. 

The program emphasizes diversity, actively seeking to include minority and female students, which is crucial for expanding access to STEM education and addressing underrepresentation in technology fields.

\subsection{Program Structure and Location}
The camps are hosted on-campus during the summer, providing students with an immersive university experience. This setting allows participants to not only develop technical skills but also engage with the academic community, exploring possibilities in higher education and STEM fields. The in-person environment fosters interactions with peers, faculty, and mentors that online formats cannot replicate. A typical day at the camp begins with a morning mini-activity designed to engage students in a specific coding challenge or data science concept, setting the tone for the day. For example, a morning session might focus on developing an algorithm for sorting data, which students can then apply throughout the day's activities. Following the initial activity, students spend time in self-paced learning sessions, where they can explore coding topics at their own pace with mentors available to answer questions and provide guidance. Midday offers a break from the intensive learning with an hour-long lunch at the campus’s all-you-can-eat cafeteria, providing an opportunity for students to socialize and relax. After lunch, students have another hour of recess at the campus recreational center, where they can unwind and participate in various physical activities, helping to balance the academic rigor with recreational fun. The afternoon includes another mini-activity, perhaps focusing on a real-world application of machine learning, followed by more self-paced learning time. A key feature of the day is a visit from a faculty member or PhD student who shares insights from their research, offering students a glimpse into advanced topics and potential career paths in STEM. This interaction not only enriches the learning experience but also provides inspiration and a sense of what is possible with the skills they are developing. The day concludes with additional self-directed learning, where students can revisit challenges, work on projects, or engage in collaborative problem-solving with their peers. Throughout the day, undergraduate mentors offer both technical support and personal guidance, while returning high school participants serve as mentors, fostering a strong support network and sense of community within the camp. This structured yet flexible schedule ensures that students receive a comprehensive educational experience, blending hands-on learning, academic exploration, and social interaction in a way that is uniquely facilitated by the on-campus environment.

\subsection{Mentorship and Long-Term Engagement}
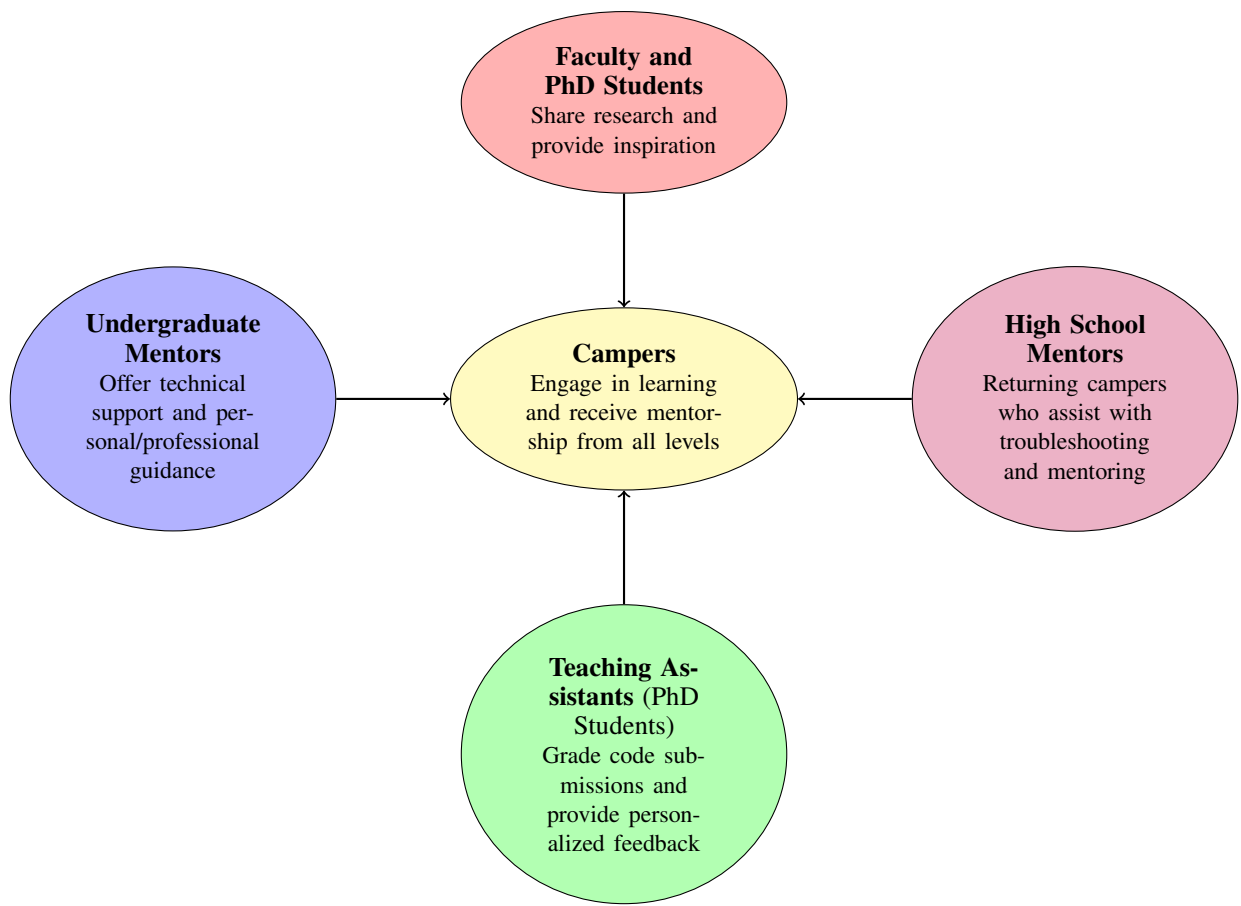
\begin{figure}[h!]
    \centering
    \resizebox{\textwidth}{!}{
    \begin{tikzpicture}[
        node distance=1cm and 1cm,
        mentor/.style={ellipse, draw=black, text centered, text width=2.8cm, minimum height=1.5cm},
        camper/.style={ellipse, draw=black, fill=yellow!30, text centered, text width=3cm, minimum height=1.5cm},
        faculty/.style={mentor, fill=red!30},
        ta/.style={mentor, fill=green!30},
        undergrad/.style={mentor, fill=blue!30},
        hsmentor/.style={mentor, fill=purple!30}
        ]

        \node (campers) [camper] {\textbf{Campers}\\ \small{Engage in learning and receive mentorship from all levels}};

        \node (faculty) [faculty, above=of campers, yshift=0.5cm] {\textbf{Faculty and PhD Students}\\ \small{Share research and provide inspiration}};
        \node (ta) [ta, below=of campers, yshift=-0.5cm] {\textbf{Teaching Assistants} (PhD Students)\\ \small{Grade code submissions and provide personalized feedback}};
        \node (undergrads) [undergrad, left=of campers, xshift=-0.5cm] {\textbf{Undergraduate Mentors}\\ \small{Offer technical support and personal/professional guidance}};
        \node (hsmentors) [hsmentor, right=of campers, xshift=0.5cm] {\textbf{High School Mentors}\\ \small{Returning campers who assist with troubleshooting and mentoring}};

        \draw[->, thick] (faculty) -- (campers);
        \draw[->, thick] (ta) -- (campers);
        \draw[->, thick] (undergrads) -- (campers);
        \draw[->, thick] (hsmentors) -- (campers);

    \end{tikzpicture}
    }
    \caption{Mentorship Structure in the Camp Program}\label{Fig.Mentor}
    \label{fig:mentorship_structure}
\end{figure}

A key feature of the camps is the opportunity for students to return as mentors, either to progress to the next level or support new participants. This mentorship system reinforces skills, builds community, and fosters leadership. Returning students gain valuable experience in teaching and collaboration, deepening their engagement in the field. The structure of this mentorship, involving previous participants, undergraduates, graduates, and faculty, is illustrated in Figure \ref{Fig.Mentor}.

\section{Student Experience}\label{Sec:Exp}
At the conclusion of each camp week (or level), participants were randomly selected for interviews to gain insights into their experiences at the data science summer camp. The interviews focused on the students' coding skills before and after the camp, their favorite aspects, challenges faced, and suggestions for improvement. For the purpose of this study we concentrate on interviews conducted on Level 3 participants as these participants have gained the knowledge and skills of all three levels of the camps. 

By the end of Level 3, interviews with three students revealed common themes such as skill development, rewarding challenges, and the value of collaboration. The students had varied coding backgrounds, from minimal prior experience to substantial self-study. One student, who had a strong background in coding, rated their experience highly, emphasizing the satisfaction of overcoming challenging labs. They stated, ``My favorite part of the camp is getting the labs done, especially the more difficult ones, which are really rewarding for me to solve." They also noted increased confidence:``I feel really confident, especially with this camp giving me practical knowledge that I hadn't had before."

Another student, with little prior coding experience, found the camp transformative. They mentioned,``Starting off, before I came to the camp, I had a negative experience with coding, but during Level 2 they started with the basics, and we proceeded from there. It definitely helped develop my confidence and made me enjoy coding." The third student, who had intermediate coding skills, found later labs and sentiment analysis particularly challenging. They highlighted the collaborative environment as key to overcoming difficulties: ``Coding is very collaborative. It’s nice to work with people on problems and help others solve problems." This appreciation for group work was echoed by others, who valued the social aspects of the camp.

\subsubsection*{Interest in AI and Data Science}
A survey conducted at the end of Level 3 revealed that 68.6\% of students expressed interest in pursuing careers in AI and data science. Of these, 35\% were drawn to AI's potential to revolutionize industries, while 20\% were intrigued by data science applications. However, 17\% were unsure about pursuing these fields, and 14\% were not interested, citing other career goals.

\subsubsection*{Key Learnings and Interests}
When asked about the most interesting aspects of AI, students frequently mentioned the diverse applications discussed during the camp. Topics such as ethical AI, autonomous vehicles, AI in engineering, and machine learning for medical imaging were particularly memorable. According to the survey, 40\% of students were most impressed by general AI and machine learning, 25\% were intrigued by the ethical implications, and 20\% found the technical aspects, like pattern recognition and decision-making speed, most fascinating. Machine learning emerged as the most popular topic, with 37.1\% of students expressing strong interest due to its real-world applications. Data science was the favorite for 28.6\% of students, appreciated for its wide-ranging applications across various industries. General AI captivated 22.9\%, and 11.4\% expressed equal interest in all topics, recognizing their collective significance for the future of technology.

\section{Conclusion and Lessons Learned}\label{SecLessons}

Throughout the camps, key lessons emerged that contributed to the program’s success. Opportunities for interaction outside the classroom, such as during recess, were essential for building camaraderie and fostering collaboration. These interactions, along with diverse mentors, helped participants from underrepresented minorities envision themselves in data science fields. Facilitating collaboration required intentional actions, like grouping students by completion levels or arranging computers to encourage teamwork. Although personalized feedback took time, it was clear that students benefited from understanding and correcting their mistakes. The impact of the camps on participants’ confidence in coding was particularly evident during the parent showcase, where students proudly explained the models they had learned, demonstrating their skills and newfound self-assurance. The summer camps provided high school students with foundational knowledge in coding, data science, and deep learning. The structured progression allowed students to build skills at a comfortable pace, while diverse mentoring ensured they felt supported and inspired. The camps also addressed disparities in STEM education, especially for underrepresented minorities. By fostering a collaborative and inclusive environment, the program significantly boosted students' technical skills and confidence. The positive outcomes—improved coding proficiency, increased interest in AI and data science careers, and heightened enthusiasm for learning—underscore the importance of such initiatives in shaping the future of technology education. The insights gained will be crucial for refining and expanding similar programs, ensuring more students benefit from these transformative experiences.
\section*{Acknowledgments}
This research was supported in part by the National Science Foundation (NSF) under award 2112533. The Seth Bonder Data Science Camps were made possible thanks to the financial support of the Seth Bonder Foundation. The development of materials and modules was led by Professor Pascal Van Hentenryck. Coordination assistance was provided by the Center for Education Integrating Science, Mathematics, and Computing (CEISMC). Special thanks go to Sirocus Barnes and Eric Koonce. The success of the 2024 Seth Bonder Data Science Camps was due to the onsite efforts of undergraduate TAs, high school mentors, graduate TAs, and coordinators. We extend a special mention to Shayna Coffsky, Iris Chen, Nethan Nagendran, Tsion Fitsum, Alexandria Sweeny, Jorge Huertas, Michael Klamkin, Stefan Faulkner, Kevin Wu, and Thomas Bruys. Lastly, coordination support was also provided by Kevin Dalmeijer and Tuba Ketenci. Making a difference requires the dedication of a committed team, and we are grateful to everyone that helped bring this project to life.

\bibliography{citation.bib}

\end{document}